\documentstyle[aps,epsf]{revtex}

\global\firstfigfalse
\global\firsttabfalse

\begin{document}

\baselineskip 12pt
\title{
\begin{flushright}
{\large UCTP-108-99}\\
{\large PRINCETON/HEP/99-3}\\
\end{flushright}
\vskip 0.15in
Development of a Straw Tube Chamber with Pickup-Pad Readout}

\author{C.~Leonidopoulos and C.~Lu}
\address{Department of Physics, 
            Princeton University, Princeton, NJ 08544}

\author{A.~J.~Schwartz} 
\address{Department of Physics, 
            University of Cincinnati, Cincinnati, OH 45221}

\maketitle              

\begin{abstract}
We have developed a straw tube chamber with pickup-pad readout. 
The mechanism for signal pickup, the size of the pickup signal, and 
the distribution of signals among neighboring pads are discussed. 
We have tested a prototype chamber in a beamtest at Brookhaven National
Laboratory and have measured chamber efficiencies in excess of 99\%. 
\end{abstract}

\section{Introduction}

In this paper we describe the construction and performance of a straw 
tube chamber with a unique type of readout: rather than signals being 
read from anode wires, signals are read from pads placed against the 
straw tubes which pick up signals induced by the gas avalanches. This 
readout scheme has three advantages: 
{\it (1)}\ fully-correlated $x$ and $y$ position information;
{\it (2)}\ the pad sizes can be chosen almost arbitrarily, for example,
          to give equal pad occupancy over a region in which 
	  particle flux is varying; and 
{\it (3)}\ the high-voltage and front-end electronics are decoupled.
For the gas avalanche signals to be picked up by the pads, the straw 
tube material must be resistive such that the straw tube wall 
does not substantially shield the signal. There are several
issues which determine the feasibility of using such a detector under 
experimental conditions: the size of the pickup signal, the level of 
noise (both capacitive and pickup), and the distribution of pickup 
signals among neighboring pads. 
Multiwire proportional chambers with resistive cathodes and pickup pad 
readout have been studied by Battistoni {\it et al.}\cite{bib:Battistoni}. 
Resistive straw tubes with pickup {\it wires\/} wrapped around the 
straws have been studied by Bychkov {\it et al.}\cite{bib:Bychkov}. 

Our final objective is to use such a chamber to trigger on tracks with 
large transverse momentum ($p^{}_T$) in the {\it HERA-B\/} experiment at 
DESY\cite{bib:hbprop,bib:hbtdr}. To accomplish this, we position three 
chamber stations within the {\it HERA-B\/} spectrometer magnet and input 
the pad hit pattern to fast coincidence logic. The logic flags hit
patterns that are characteristic of tracks which bend little within the 
magnetic field and which lie at larger angles with respect to the beam.
To minimize hit multiplicities and simplify the trigger logic, we desire 
that the pickup signal be as localized as possible, preferentially 
occurring on only one pad.

\section{Chamber Construction}

\subsection{Overview}

\begin{figure}[ht]	
\centerline{\epsfxsize 4.0 truein \epsfbox{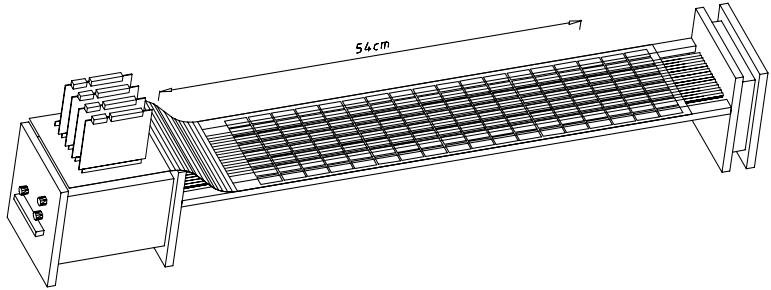}}   
\vskip .2 cm
{\caption[Short caption for table of contents ]
{\label{fig:prototype} \small Stereo view of prototype chamber~A. 
The pickup pads are on the downward side of the Kapton circuit board, 
but are shown (along with the straw tubes) for clarity. The front-end 
preamplifier cards are at the left end. }}
\end{figure}

We have constructed two prototype chambers which we refer to as 
chamber~A and chamber~B. The chambers are similar and consist of 
two distinct parts:
{\it (1)}\ the straw tubes, their endplates, and 
the high voltage distribution cards; and 
{\it (2)}\ the pickup pads and their preamplifier cards.
The straw tubes have a diameter of 5.0~mm and are configured in two 
rows of sixteen straw tubes each. For chamber~A, the tubes have a length 
of 60~cm and the rows are offset with respect to each other by one straw 
tube radius. This ``close-packed'' geometry is especially resistant 
to mechanical stresses. For chamber~B, the tubes have a length of 
40~cm and the rows are offset by a smaller amount, only 0.6~mm. 
Chamber~B also has a thin line of conductive paint applied to the 
surface of every third straw tube. This paint is held at ground
potential, and its purpose is to limit the pickup signal to a single 
pad (to be discussed later). In all other aspects the two 
chambers are identical. The pickup pads for both chambers have 
dimensions $1.5\times3.0$~cm$^2$ and are etched on one side of a 
printed circuit board. 
The arrangement of straw tubes and pad plane is shown 
in Fig.~\ref{fig:prototype}. 

\subsection{Straw Tubes}

The straw tubes are rolled\cite{bib:Stone} out of two-layer material: 
the inner layer consists of carbon-loaded Kapton with a resistivity of 
approximately 1~M$\Omega$/square\cite{bib:Dupont}, and the outer layer 
consists of Mylar. The Kapton layer serves as the cathode and is held 
at ground potential, while the Mylar layer helps provide a gas seal. 
The thickness of the Kapton is 0.001 inch, and the thickness of the 
Mylar is 0.0005 inch. The straws are strung with gold-plated tungsten 
wire having a diameter of 20~$\mu$m. 
For most of our tests we 
used a gas mixture of 90\% Argon and 10\% Methane (``P10'').

\subsection{Pickup Pads}

The pad plane consists of a two-layer printed circuit 
board with copper pads on one side and traces on the other. 
The traces connect to the pads via plated through-holes and 
carry signals to one end of the pad plane, where preamplifier 
cards are mounted. To minimize detector mass, the material used 
for the printed circuit board is Kapton. The thickness of the Kapton 
is only 0.006~inches (including the Cu pads and traces), which 
corresponds to 0.28\% of a radiation length. 
The trace width is 0.006 inches. The front-end readout cards are 
based upon the ASD-8 amplifier-shaper-discriminator ASIC developed 
at the University of Pennsylvania\cite{bib:asd8}.  

\section{Bench Measurements and Simulation Results}

\begin{figure}[ht]	
\centerline{\epsfxsize 2.5 truein \epsfbox{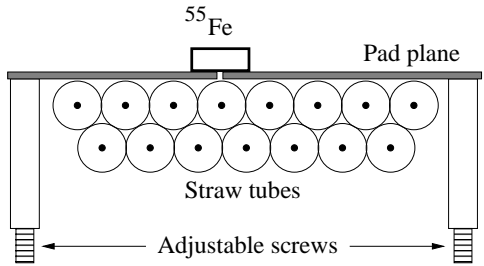}}   
\vskip .2 cm
{\caption[Short caption for table of contents ]
{\label{fig:test_setup} \small The test setup used to measure the 
pickup signal size 
and the distribution of pickup signals among neighboring pads. }}
\end{figure}

We have conducted a series of bench tests to measure the signal size, 
noise, and the relative sizes of pickup signals among neighboring 
pads. The test setup used is shown in Fig.~\ref{fig:test_setup}.

\subsection{Signal Size}

To measure signal size, we positioned an $^{55}$Fe source 
($E^{}_\gamma =5.9$~KeV) above a small hole drilled in the 
pad plane. This hole is located directly above a straw tube 
as shown in Fig.~\ref{fig:test_setup}. 
We measured the signal collected on both the anode 
wire and the pickup pad using an Ortec multichannel analyzer (MCA) 
system consisting of a 142PC charge amplifier, a 570 spectroscopy 
amplifier, and a 419 MCA PC plug-in board.  The spectra show a 
significant signal from the pickup pad: the ratio of pad signal 
to anode wire signal is about 1/3. 

\begin{figure}[ht]	
\centerline{\epsfxsize 3.2 truein \epsfbox{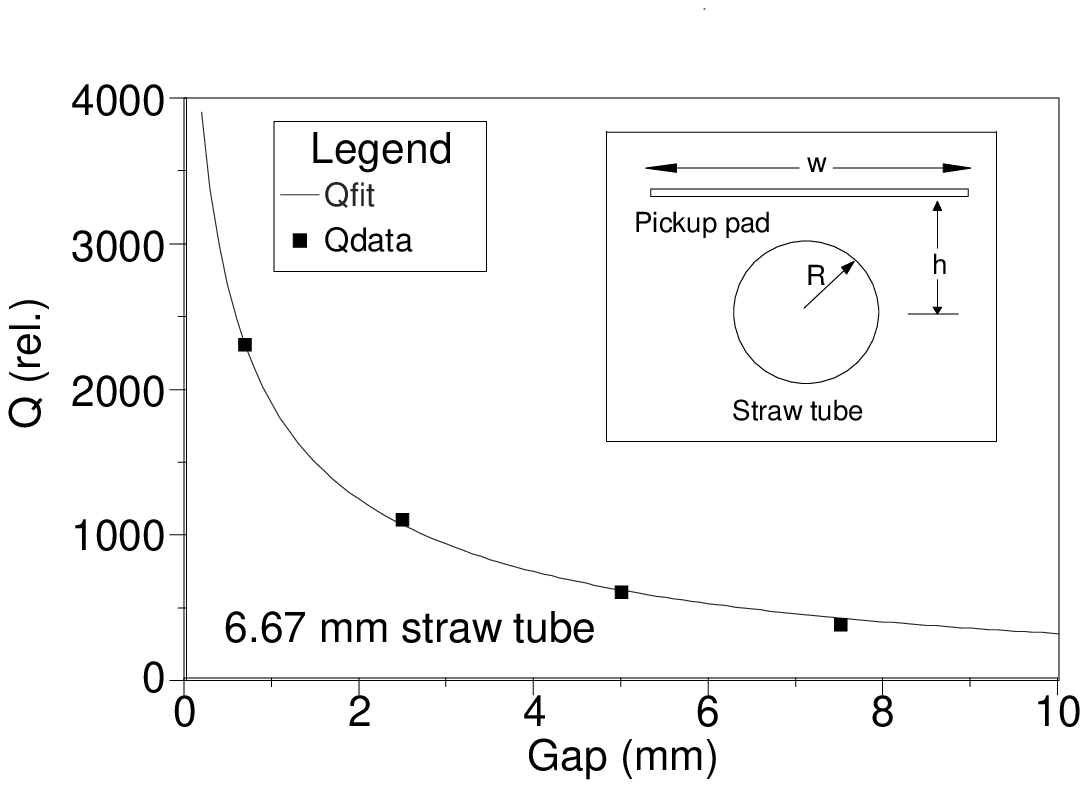}}      
\vskip .2 cm
{\caption[Short caption for table of contents ]
{\label{fig:sig_vs_gap} \small The dependence of the pickup signal 
on the gap between the straw tubes and the pickup pads. The straw
tubes used for these measurements have a diameter of 6.67~mm. }}
\end{figure}

The size of the pickup signal is very sensitive to the size of the 
gap between the straw tubes and the pad plane. We have measured this 
dependence using the same test setup; the resulting data is shown in 
Fig.~\ref{fig:sig_vs_gap}. 
We account for this data  quantitatively by considering the geometry 
of the insert in Fig.~\ref{fig:sig_vs_gap}. For a single straw tube 
and an infinitely long pickup strip oriented parallel to it, the 
capacitance per unit length between the straw tube and pickup strip 
is\cite{bib:Kraus}:
\begin{equation}
C = \frac{4\epsilon^{}_0\tan^{-1}(w/2s)}{\ln k},
\label{eqn:sig_vs_gap}
\end{equation}
where 
$\epsilon^{}_0$ is the permittivity of free space (neglecting the
effect of the thin Mylar layer of the straw tube wall), 
$w$ is the width of the pickup strip,
$k=h/R + \sqrt{(h/R)^2-1}$, and
$s=h-R/k$. In these expressions $h$ is the distance between
the center of the tube and the pickup strip, and $R$ is the 
radius of the tube.
If the potential on the straw tube is $V$, then the signal on the 
pickup strip per unit length is $CV$. It is due to the factor 
$C$ that the pickup signal depends strongly on the ratio $h/R$. 
The signal size predicted by Eq.~(\ref{eqn:sig_vs_gap}) is 
superimposed on the data in Fig.~\ref{fig:sig_vs_gap} and 
shows very good agreement.

The basic assumption of this model is that the straw tube wall is 
an equipotential surface; i.e., the charge induced on the tube wall
has sufficient mobility to form an equipotential surface within a 
time scale much less than the response time of the readout 
electronics. This time scale is expected to be of order 
$\epsilon/\sigma$\cite{bib:Bochove}, where $\epsilon$ is the 
permittivity of Kapton and $\sigma$ is its bulk conductivity. 
Since our film has a resistivity of 1~M$\Omega$/square and 
a thickness~$T$ of 0.001 inch, 
$\sigma = 1/\rho = 1/(R\cdot T) = (25.4~\Omega$-m)$^{-1}$. 
If we take the permittivity of Kapton to be that of free space 
$(\epsilon_0)$, then the time scale required is 225~ps. This 
interval is relatively short: the fraction of signal charge 
collected in time $t$ is 
$(2\xi)^{-1}\ln [1+ 2\mu^{}_+V^{}_A t/(a^2\xi)]$, where
$\xi\equiv \ln (b/a)$,  
$a$ is the radius of the anode wire,
$b$ is the radius of the straw tube,
$\mu^{}_+$ is the mobility of the positive ions,
and $V^{}_A$ is the anode wire voltage.
Inserting values $a=10$~$\mu$m, $b=2.5$~mm, 
$\mu^{}_+\approx 150$~mm$^2/(V$-sec), and 
$V^{}_A\approx 1350$~V, we find that in 225~ps only 
1.4\% of the signal charge has been collected. 

\subsection{Distribution of Pickup Signals}

\begin{figure}[htb]	
\centerline{\epsfxsize 3.4 truein \epsfbox{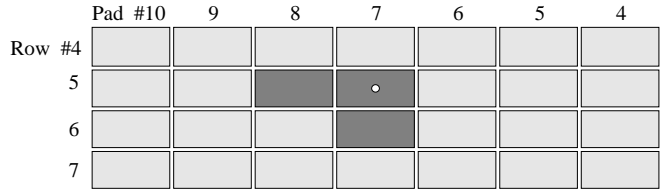}}     
\vskip .2 cm
{\caption[Short caption for table of contents ]
{\label{fig:pad_geometry} \small A section of the pad plane used to 
study the distribution of pickup signals among neighboring pads. The 
$^{55}$Fe source is positioned above pad \#7, row \#5. There is 
a small hole drilled in the center of this pad to allow X-rays 
from the source to enter the straw tube below. }}
\end{figure}

We have measured the distribution of pickup
signals among neighboring pads using an $^{55}$Fe source.
The pad geometry studied is shown in Fig.~\ref{fig:pad_geometry}. 
The $^{55}$Fe source is placed above pad \#7, row \#5. There is a 
small hole drilled in the center of this pad to allow X-rays from the 
source to enter the straw tube below.  We trigger on signals from this 
pad and measure the signals picked up by neighboring pads. The results 
are summarized in Table~\ref{tab:spread}.

\subsubsection{Distribution of Signals Along the Straw Tube Direction}

From Table~\ref{tab:spread} we observe that the signal distribution
along the straw tube direction (pad \#8, row \#5) is narrow. There are 
two mechanisms affecting the distribution in this direction:
{\it (a)\,} the direct signal-induction mechanism, and
{\it (b)\,} signal propagation along the transmission line formed by 
the anode wire and the cathode tube. The signal distribution due to 
$(a)$ can be calculated analytically using Green's reciprocation 
theorem\cite{bib:Greens}; a solution is given in the long version 
of this paper\cite{bib:Leonido}. The resulting charge distribution 
on the pads after 6~ns (the shaping time of the ASD-8) 
is relatively narrow: the standard deviation of a Gaussian fit to the 
distribution is only $\sim$\,1.4 mm. This value is consistent with
the value measured (Table~\ref{tab:spread}); however, the measurement 
is dominated by noise and also receives contributions from wider angle 
photons that produce gas avalanches closer to the pad boundary. Thus, 
we interpret the values in Table~\ref{tab:spread} as upper bounds.

The signal distribution due to propagation via the transmission
line formed by the anode wire and the cathode tube is also calculated
in Ref.~\cite{bib:Leonido}. Since the tube wall is made 
of a highly-resistive material, the signal propagates with strong 
attenuation. The calculation predicts that for a straw tube
resistivity of 1~M$\Omega$/square, the signal is attenuated by 
a factor of 89 over a distance of 1~cm. We thus conclude that 
for pad lengths 
$\mathrel{\rlap{\raise 0.511ex \hbox{$>$}}
               {\lower 0.511ex \hbox{$\sim$}}} 3$~cm, the signal 
distribution among pads due to this mechanism is also narrow.

\subsubsection{Distribution of Signals Perpendicular to the
               Straw Tube Direction}

Table~\ref{tab:spread} shows that the pickup signal on the pads 
transverse to the straw tube direction is 12\% of that on the main 
pad. This result has been successfully simulated using two models: 
{\it (1)\/}\ a simple ``capacitive coupling'' model, and 
{\it (2)\/}\ a finite element analysis (FEA) simulation.

\begin{figure}[ht]	
\centerline{\epsfxsize 5.0 truein \epsfbox{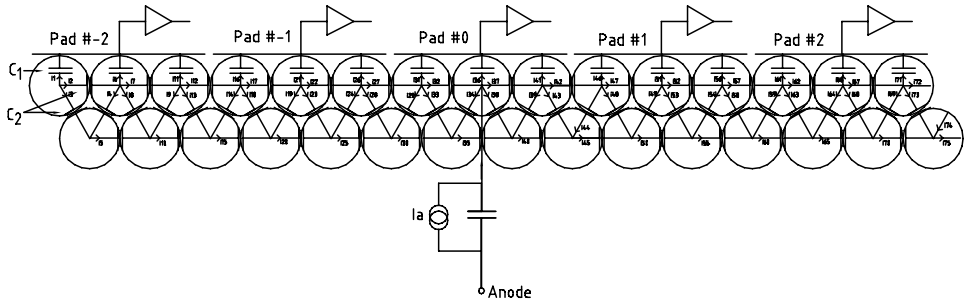}}   
\vskip .2 cm
{\caption[Short caption for table of contents ]
{\label{fig:ccmodel} 
\small The capacitive coupling model for prototype chamber~A.
This model is used to calculate the distribution of pickup signals
perpendicular to the straw tube direction. }}
\end{figure}

The capacitive coupling model corresponding to chamber~A is shown 
in Fig.~\ref{fig:ccmodel}. In this figure, $C1$ is the capacitance
between a straw tube and a pad, and $C2$ is the capacitance between
adjacent straw tubes. Note that one pad (${\rm width}=1.5$~cm) covers 
three straw tubes (${\rm diameter}=5$~mm). The gas avalanche is modeled 
as a current source injecting charge $Q$ into a single node as 
indicated in the figure. We apply Kirchhoff's law to this circuit 
to obtain a set of linear, coupled equations. We then solve these 
equations numerically to obtain the charge induced on each pad. We
have calculated the induced charge for two cases: 
$C1/C2=1$ and $C1/C2=2$. The first case corresponds to a gap size 
of approximately 0.1~mm, 
while the latter case corresponds to a gap size of only 
$\sim$\,0.012~mm. The result for $C1/C2=2$ is very close to 
the measured value: the ratio of the signal on an adjacent pad 
to that on the main pad is~0.13, while the measured value is 0.12 
(see Table~\ref{tab:spread}). 
The result obtained with the FEA simulation is identical.
A detailed description of the FEA calculation is given in 
Ref.~\cite{bib:Leonido}.

\subsubsection{Limiting the Pickup Signal}

\begin{figure}[ht]	
\centerline{\epsfxsize 2.4 truein \epsfbox{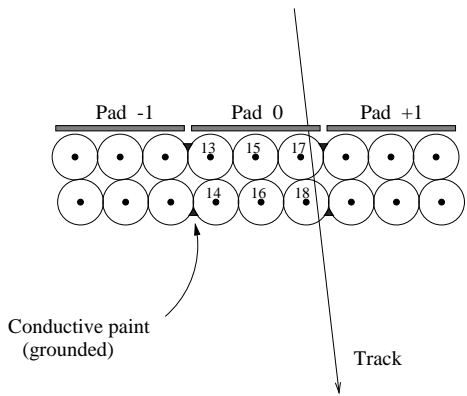}}      
\vskip .2 cm
{\caption[Short caption for table of contents ]
{\label{fig:silver_paint} 
\small To limit the distribution of pickup signals among
neighboring pads, a thin line of conductive paint is applied 
to the juncture between straw tubes which lie at pad boundaries. 
The paint is held at ground potential via contact with the chamber
endplates. }}
\end{figure}

Our bench measurements and simulation results indicate that the 
distribution of pickup signals perpendicular to the direction of the
straw tubes ---\,when fit to a Gaussian function\,--- has a standard 
deviation of $\sim$\,0.5~pad.\footnote{This spread is for a straw tube 
diameter of 5.0~mm and a pad width of 15~mm.} This spread is relatively 
large for triggering purposes. To prevent adjacent pads from recording 
hits, we have 
developed a technique to limit the pickup signal to a single pad. This 
technique is as follows: at the juncture between straw tubes which lie 
at pad boundaries, we apply a thin line of conductive paint as shown in 
Fig.~\ref{fig:silver_paint}. This paint is held at ground potential via 
contact with the chamber endplates and provides a preferred location for 
electric field lines to end. In this manner the capacitive coupling 
between adjacent straw tubes is reduced. This mechanism has been 
simulated by the FEA program, and the resultant distribution of 
pickup signals 
indicates that the ratio of signal on an adjacent pad to that on 
the main pad is only 1/62. We have measured this ratio using the 
pad arrangement of Fig.~\ref{fig:pad_geometry} and obtained~1/38.
This value is in rough agreement with that calculated and is 
significantly smaller than that measured without the conductive 
paint applied, $\sim$\,1/8 (Table~\ref{tab:spread}). 

\section{Beamtest Results}

\begin{figure}	
\centerline{\epsfxsize 4.5 truein \epsfbox{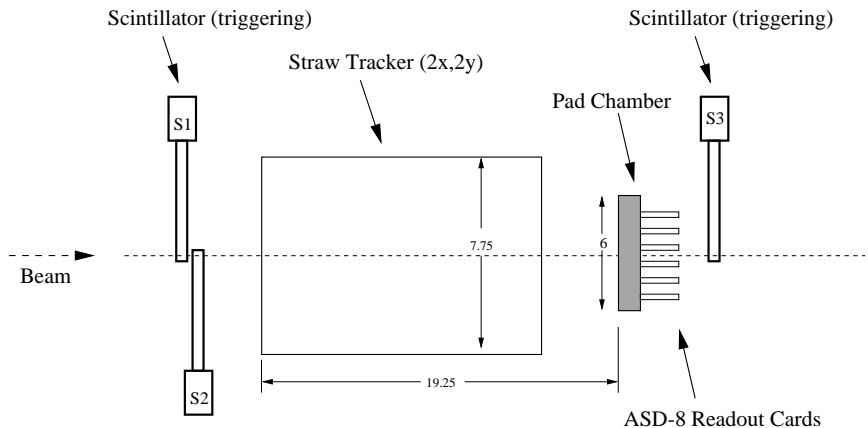}}      
\vskip .2 cm
{\caption[Short caption for table of contents ]
{\label{fig:bnl_plan_view} 
\small Plan view of the apparatus used for the beamtest
of prototype chamber~A at BNL. The pad chamber is oriented
vertically, out of the page. All dimensions listed are in inches. }}
\end{figure}

We have tested prototype chamber~A in a beamtest at Brookhaven National 
Laboratory (BNL) in order to measure the efficiency and noise of the 
chamber under experimental conditions. The beamtest was conducted at 
the Alternating Gradient Synchrotron (AGS) using a secondary beam 
consisting mostly of pions with momentum in the range $3-8$~GeV/$c$. A 
plan view of the apparatus used is shown in Fig.~\ref{fig:bnl_plan_view}. 
Immediately upstream of the prototype chamber was a conventional 
straw tube tracking chamber consisting of two $x$-measuring and two 
$y$-measuring planes. A LeCroy~4448 latch 
module was used to record hits from the pad chamber, while several 
LeCroy~2277 TDC's were used to record hits from the straw tube tracker. 
The TDC hit information allowed us to reconstruct tracks, project them 
to pads of the prototype chamber, and measure the efficiency of the 
chamber and the distribution of pickup signals among neighboring pads. 

\subsection{Chamber Efficiency}

The efficiency of the pad chamber was measured by reconstructing tracks in 
the tracker, projecting them to the pad chamber, and recording whether the 
pad(s) to which a track projects has a signal. The results are listed in 
Table~\ref{tab:beamtest_eff}. For high voltage (HV) applied to both layers 
of straw tubes, the efficiency is at least 99.5\% for all three discriminator 
threshold settings. The table also shows that when high voltage is applied 
to (at least) the upper layer of straws, there is very little dependence 
of the efficiency on the discriminator threshold setting. This indicates 
that even our highest threshold setting (1.75~V) corresponds to a signal 
well below typical pulse-heights.

\subsection{Distribution of Pickup Signals}

\begin{figure}[ht]	
\hbox{\hspace*{0.80in}
\mbox{\epsfxsize 2.20 truein \epsfbox{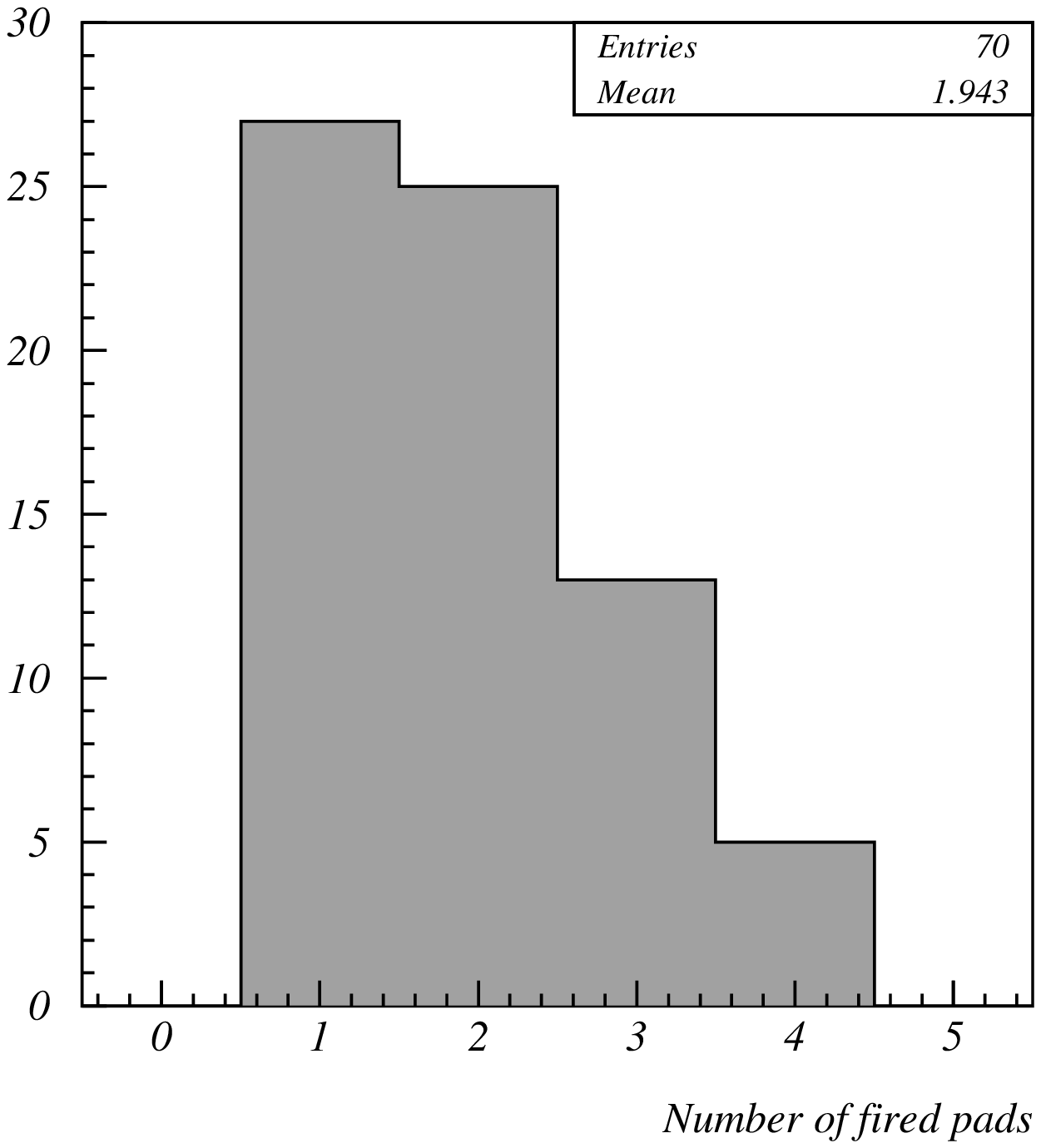}}   
\hspace*{0.80in}
\mbox{\epsfxsize 2.20 truein \epsfbox{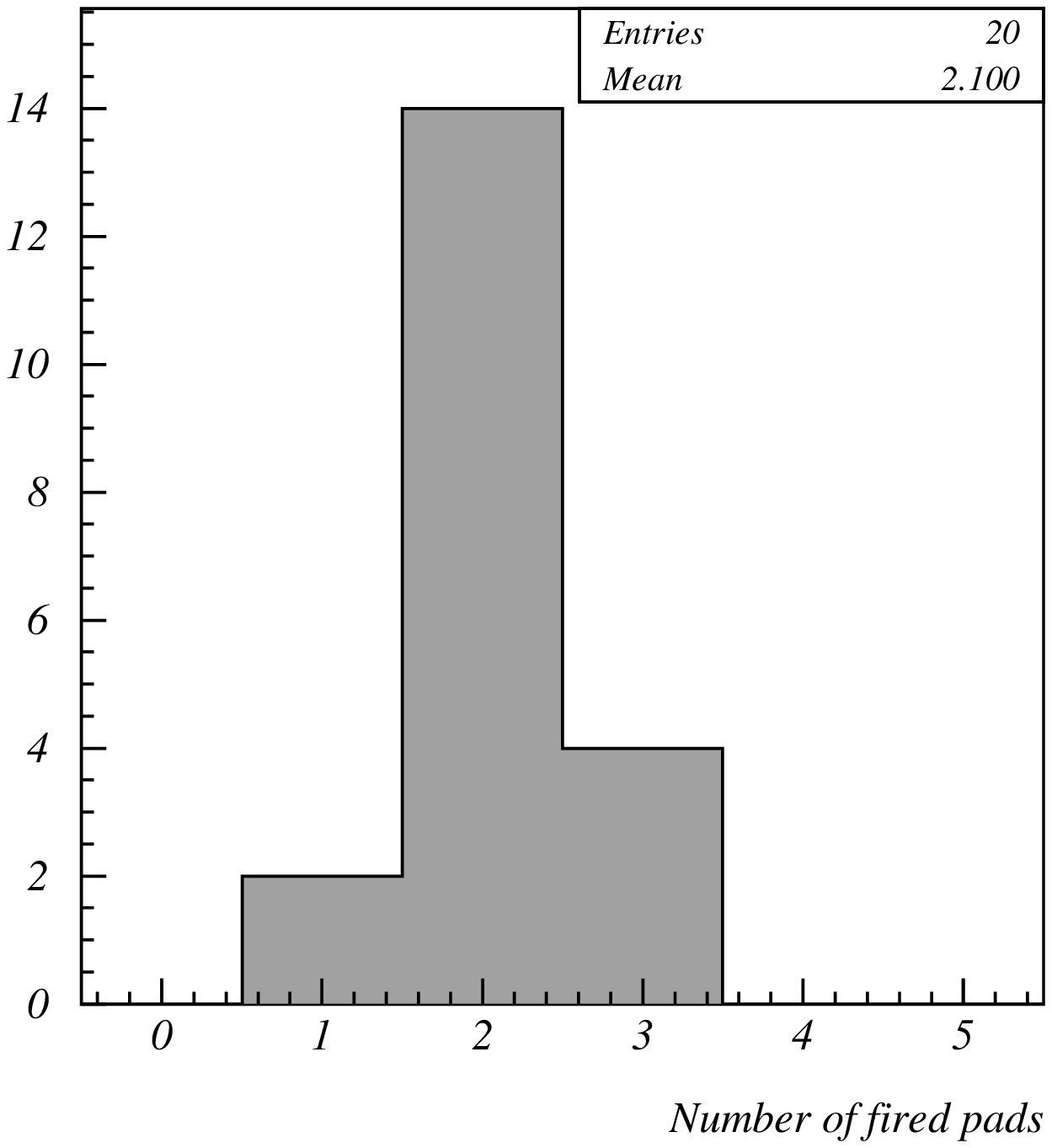}}   
}
\vskip .2 cm
{\caption[Short caption for table of contents ]
{\label{fig:beamtest_spread} 
\small The number of contiguous pads in $x$ (perpendicular 
to the straw tubes) which had hits due to a track passing 
through a double-layer of straws.
{\it (a)}\ The track projects to within 2~mm of the center of a pad.
{\it (b)}\ The track projects to within 2~mm of the center in the 
           $y$-coordinate, and to within 2~mm of the {\it edge} in 
           the $x$-coordinate. }}
\end{figure}

We measured the distribution of pickup signals among neighboring pads 
by projecting reconstructed tracks to the pad chamber and recording 
the total number of pads which had hits; i.e., including those pads 
adjacent to the pad to which the track projects. We implement this 
procedure for two subsets of tracks: those which project to within 
2~mm of the center of a pad, and those which project to within 2~mm 
of the center in the $y$-coordinate and to within 2~mm of the 
{\it edge} in the $x$-coordinate. We expect the latter subset
--\,on average\,-- to have pickup signals distributed over a 
larger number of pads. For both subsets the number of contiguous 
pads which had hits is shown in Fig.~\ref{fig:beamtest_spread}. 
The mean number of pads with 
hits is 1.9 for tracks which project to the center of a pad, and 
2.1 for tracks which project to the edge of a pad. 
Fig.~\ref{fig:beamtest_spread} shows that even for the optimal case of 
a gas avalanche occurring in the center of a pad, the frequency of only 
one pad having a hit is only 39\%. To increase this fraction, we 
developed the technique of applying conductive paint to every third
straw tube as discussed previously.

\section{Acknowledgements}

We gratefully acknowledge the strong support of the BNL AGS testbeam 
group, in particular Alan Carroll and Ed Schwaner. We are indebted to 
William Sands and Robert Klemmer of the Princeton Experimental 
Particles Laboratory for their excellent work in constructing the 
prototype chambers, and to Stanley Chidzik of the electronics shop 
for producing the ASD-8-based preamplifier cards. Finally, we thank 
Mitch Newcomer,  Frank Shoemaker, and Kirk McDonald for many useful 
discussions. 

\begin{table}
\caption{\label{tab:spread}  
The signals picked up by three neighboring pads when triggering on 
pad \#7, row \#5. The first and second pads listed are located along
the straw tube direction, and the first and third pads listed are 
located perpendicular to the straw tube direction.}
\begin{tabular}{l|ccc}
  &  {\bf {\boldmath Pad \#7, Row \#5}}
  &  {\bf {\boldmath Pad \#8, Row \#5}}
  &  {\bf {\boldmath Pad \#7, Row \#6}} \\
\hline
Signal (mV)           &  207  &  $<15$    &  $\sim 26$    \\
Relative size  &  1.0  &  $<0.07$  &  $\sim 0.12$  \\
\end{tabular}
\end{table}

\begin{table}
\caption{\label{tab:beamtest_eff}  
The efficiency of prototype chamber~A as measured in a beamtest 
at BNL, for three different HV configurations and three different 
ASD-8 discriminator thresholds.}
\begin{tabular} {r|ccc}
  &  \multicolumn{3}{c}{\bf {\boldmath ASD-8 Threshold}} \\
  &  {\bf Low (1.51 V)} & {\bf Medium (1.63 V)} & {\bf High (1.75 V)} \\
\hline
HV bottom layer  &  0.913  &  0.905  &  0.887  \\
HV top    layer  &  0.953  &  0.949  &  0.945  \\
HV both layers   &  0.998  &  0.997  &  0.995  \\
\end{tabular}
\end{table}


\begin{references}

\bibitem{bib:Battistoni}
		G.\ Battistoni {\it et al.},
                     Nucl.\ Instr. Meth.\ {\bf 176}, 297 (1980);
                     Nucl.\ Instr. Meth.\ {\bf 152}, 423 (1978). 

\bibitem{bib:Bychkov} 
		V.\ N.\ Bychkov {\it et al.}, 
                     Nucl.\ Instr.\ Meth.\ {\bf A367}, 276 (1995). 

\bibitem{bib:hbprop} 
		T.\ Lohse {\it et al.}, {\it HERA-B, An Experiment 
                 to Study CP Violation in the B System Using an Internal 
                 Target at the HERA Proton Ring\/} (Proposal), 
                 DESY-PRC 94/02 (1994).  

\bibitem{bib:hbtdr} 
		E.\ Hartouni {\it et al.}, {\it HERA-B, An Experiment 
                 to Study CP Violation in the B System Using an Internal 
                 Target at the HERA Proton Ring\/} (Technical Design Report), 
                 DESY-PRC 95/01 (1995).  

\bibitem{bib:Stone} 
		Stone Industrial, College Park, Maryland;
                    Lamina Dielectrics Ltd., Billingshurst, UK. 

\bibitem{bib:Dupont} 
		XC100 Polyimide Film (Kapton), manufactured by
                     Dupont Films, Circleville, Ohio. 

\bibitem{bib:asd8} 
		F.\,M.\,Newcomer {\it et al.}, 
                     IEEE Trans.\ Nucl.\ Sci.\ {\bf 40}, 630 (1993).  

\bibitem{bib:Kraus} 
		John D.\ Kraus, {\it Electromagnetics, 4th Ed.\/} 
                    (McGraw-Hill, New York, 1992), Sec.~4-15. 
                    The capacitance to a strip of width $w$ is 
                    $(\pi/2)\tan^{-1}(w/2s)$ times the capacitance to
                    the entire plane in which the strip lies.

\bibitem{bib:Bochove} 
		R.\ J.\ Bochove and J.\ F.\ Walkup, 
                     Am.\ J.\ Phys.\ {\bf 58\,(2)}, 131 (1990).  

\bibitem{bib:Greens} 
		See for example: W.\ R.\ Smythe, 
                    {\it Static and Dynamic Electricity, 3rd Ed.\/}
                    (McGraw-Hill, New York, 1968), Sec.~2.12.

\bibitem{bib:Leonido} 
		C.\ Leonidopoulos, {\it et al.}, 
		{\it Development of a Straw Tube Chamber with Pickup-pad 
		Readout,\/} to appear in Nucl.\ Instr.\ Meth.~A. 


\end{references}
\end{document}